\documentclass[a4paper,fleqn,usenatbib]{mnras}
\usepackage{graphicx}

\newcommand{\vsini}{\mbox{$v \sin i$}}

\newcommand{\kmps}{km\,s$^{-1}$}

\newcommand{\sqiglt}{\hbox{\rlap{\lower.55ex \hbox {$\sim$}}
	\kern-.3em \raise.4ex \hbox{$<$}\,}}
\newcommand{\sqiggt}{\hbox{\rlap{\lower.55ex \hbox {$\sim$}}
	\kern-.3em \raise.4ex \hbox{$>$}\,}}

\title[WASP-South hot Jupiters]{WASP-South hot Jupiters: WASP-178b, WASP-184b, WASP-185b \&\ WASP-192b}

\author[Hellier et al.]{Coel Hellier$^{1}$,
D.R. Anderson$^{1,2}$, 
K. Barkaoui$^{3,4}$,
Z. Benkhaldoun$^{3}$,
F. Bouchy$^{5}$,\newauthor  
A. Burdanov$^{6}$,
A. Collier Cameron$^{7}$,  
L. Delrez$^{6,8}$,
M. Gillon$^{6}$,
E. Jehin$^{6}$, 
L.D. Nielsen$^{5}$, \newauthor    
P.F.L. Maxted$^{1}$, 
F. Pepe$^{5}$, 
D. Pollacco$^{2}$,    
F.J. Pozuelos$^{4,6}$,
D. Queloz$^{8}$, 
D. S\'egransan$^{5}$, \newauthor  
B. Smalley$^{1}$,  
A.H.M.J. Triaud$^{9}$,  
O.D. Turner$^{1,5}$, 
S. Udry$^{5}$, and 
R.G. West$^{2}$\\    
$^{1}$Astrophysics Group, Keele University, Staffordshire, ST5 5BG, UK\\
$^{2}$Department of Physics, University of Warwick, Gibbet Hill Road, Coventry CV4 7AL, UK\\
$^{3}$Oukaimeden Observatory, High Energy Physics and Astrophysics Laboratory, Cadi Ayyad University, Marrakech, Morocco\\ 
$^{4}$Astrobiology Research Unit, Universit\'e de Li\`ege, Li\'ege, Belgium\\ 
$^{5}$Observatoire astronomique de l'Universit\'e de Gen\`eve
51 ch. des Maillettes, 1290 Sauverny, Switzerland\\
$^{6}$Space sciences, Technologies and Astrophysics Research (STAR) Institute, Universit\'e de Li\`ege, Li\`ege 1, Belgium\\
$^{7}$SUPA, School of Physics and Astronomy, University of St.\ Andrews, North Haugh,  Fife, KY16 9SS, UK\\
$^{8}$Cavendish Laboratory, J J Thomson Avenue, Cambridge, CB3 0HE, UK\\
$^{9}$School of Physics \&\ Astronomy, University of Birmingham, Edgbaston, Birmingham, B15 2TT, UK
}

\begin{document}

\date{date}
\pagerange{range}

\maketitle

\begin{abstract}
We report on four new transiting hot Jupiters discovered by the WASP-South survey. WASP-178b transits a $V$ = 9.9, A1V star with $T_{\rm eff}$ = 9350 $\pm$ 150 K, the second-hottest transit host known.  It has a highly bloated radius of 1.81 $\pm$ 0.09 R$_{\rm Jup}$, in line with the known correlation between high irradiation and large size. With an estimated temperature of 2470 $\pm$ 60 K, the planet is one of the best targets for studying ultra-hot Jupiters that is visible from the Southern hemisphere. The three host stars WASP-184, WASP-185 and WASP-192 are all post-main-sequence G0 stars of ages 4--8 Gyr.  The larger stellar radii (1.3--1.7 M$_{\odot}$) mean that the transits are relatively shallow (0.7--0.9\%) even though the planets have moderately inflated radii of 1.2--1.3 R$_{\rm Jup}$.   WASP-185b has an eccentric orbit ($e$ = 0.24) and a relatively long orbital period of 9.4 d. A star that is 4.6 arcsec from WASP-185 and 4.4 mag fainter might be physically associated. 
\end{abstract}

\begin{keywords}
Planetary Systems --  stars: individual (WASP-178, WASP-184, WASP-185, WASP-192)
\end{keywords}

\section{Introduction}
Since its start in May 2006 the WASP-South survey for transiting exoplanets operated until mid 2016, obtaining data on over 2500 nights and recording 400 billion photometric data points on 10 million stars.  From 2006 to mid-2012 WASP-South used 200-mm, f/1.8 lenses, searching for transits of stars of $V$ = 9--13, and obtaining typically 20\,000 data points on each star. Coverage adds up to the whole sky between declination +8$^{\circ}$ and  --70$^{\circ}$, other than the crowded galactic plane, with each field being observed in typically 3 or 4 different years.  In mid-2012 WASP-South switched to 85-mm, f/1.2 lenses, changing the useful magnitude range to $V$ = 6.5--11.5, with the aim of finding the very brightest hot-Jupiter hosts such as WASP-189 \citep{2018arXiv180904897A}. 

WASP-South transit candidates proved well matched to follow-up with the 1.2-m Euler telescope and CORALIE spectrograph, teamed with the TRAPPIST-South photometric telescope and (more recently) TRAPPIST-North, which together observed 1600 planet candidates. So far WASP-South has led to the announcement of 154 transiting exoplanets (34 of them jointly with data from WASP-South's northern counterpart, SuperWASP).\footnote{See https://wasp-planets.net}

Follow-up of WASP-South candidates is now nearing completion, and in any case such surveys are rapidly being superseded by the space-based TESS survey \citep{2016SPIE.9904E..2BR}.   We report here four new transiting hot Jupiters. While WASP-184b and WASP-192b are routine hot Jupiters transiting fainter, $V$ = 12, stars, WASP-178b transits a bright A1V star that is the hottest of the WASP planet hosts, while WASP-185b has an eccentric, 9.4-d orbit.  

\begin{table}
\caption{Observations\protect\rule[-1.5mm]{0mm}{2mm}}  
\begin{tabular}{lcr}
\hline 
Facility & Date & Notes \\ [0.5mm] \hline
\multicolumn{3}{l}{{\bf WASP-178:}}\\  
WASP-South & 2006 May--2014 Aug & 101\,600 points \\ 
CORALIE  & 2017 Apr--2018 Jul  &   23 RVs \\
EulerCAM  & 2018 Mar 26 & {\it I} filter \\ 
\multicolumn{3}{l}{{\bf WASP-184:}}\\  
WASP-South & 2007 Feb--2012 Jul & 24\,300 points \\ 
CORALIE  & 2015 Jun--2018 Jul  &   19 RVs \\
TRAPPIST-South & 2016 Mar 05 & blue-block \\
EulerCAM  & 2018 Apr 11 02 & $R$ filter \\ 
\multicolumn{3}{l}{{\bf WASP-185:}}\\  
WASP-South & 2006 May--2012 Jun & 34\,000 points \\ 
CORALIE  & 2015 Jun--2018 Aug  &   24 RVs \\
TRAPPIST-South & 2014 Apr 09 & $z$ band \\
TRAPPIST-North & 2019 Jun 09 & $z$ band \\
\multicolumn{3}{l}{{\bf WASP-192:}}\\  
WASP-South & 2006 May--2012 Jul & 42\,200 points \\ 
CORALIE  & 2016 Jun--2019 Apr  &   12 RVs \\
TRAPPIST-South & 2016 Apr 17 & blue-block \\
TRAPPIST-South & 2019 Jun 06 & $I+z$ band \\
\end{tabular} 
\end{table}

\section{Observations}
The WASP-South photometry was accumulated into multi-year lightcurves for every catalogued star, which were then searched for transits using automated routines \citep{2006PASP..118.1407P,2007MNRAS.380.1230C}, followed by human vetting of the search outputs.  Planet candidates were then listed for followup observations by the TRAPPIST-South 0.6-m robotic photometer (e.g.~\citealt{2013A&A...552A..82G}) and the Euler/CORALIE spectrograph (e.g.~\citealt{2013A&A...551A..80T}).  Transit photometry for the stars reported here was also obtained with the EulerCAM photometer (e.g.~\citealt{2012A&A...544A..72L}) and with TRAPPIST-North \citep{2019AJ....157...43B}.  Our observations are listed in Table~1. 

For three of our stars (WASP-184, WASP-185 and WASP-192) the CORALIE spectra were reduced to radial-velocity measurements using a standard G2 mask \citep{2002A&A...388..632P}, while for the hotter star WASP-178 we used an A0 mask. The resulting values are listed in Table~A1.

As we routinely do for WASP-South planet discoveries, we used the WASP photometry, typically spanning 6 months of observation in a year and several years of coverage, to look for rotational modulations of the planet-host stars.  Our methods are detailed in \citet{2011PASP..123..547M}. For the four stars reported here we found no significant modulations with upper limits of 1--2 mmags (as reported in the Tables for each star).

\section{Spectral analyses}
We combined the CORALIE spectra for each object in order to make a spectral analysis. For three of the stars discussed here (WASP-184, WASP-185 and WASP-192) we adopt the same methods used in recent  WASP-South papers (e.g.~\citealt{2019MNRAS.482.1379H}), as described by \citet{2013MNRAS.428.3164D}. Thus we estimated the effective temperature, $T_{\rm eff}$, from the H$\alpha$ line, and the surface gravity, $\log g$, from Na~{\sc i} D and Mg~{\sc i} b lines.  We also translate the  $T_{\rm eff}$ value to give an indicative spectral type.  To estimate the metallicity, [Fe/H],  we make  equivalent-width measurements of unblended Fe~{\sc i} lines, quoting errors that take account of the uncertainties in $T_{\rm eff}$ and $\log g$. We use the same Fe~{\sc i} lines to measure $v \sin i$ values, taking into account the CORALIE instrumental resolution ($R$ = 55\,000) and adopting macroturbulence values from \citet{2014MNRAS.444.3592D}.    The spectral analysis values are reported in the Tables for each star. 

WASP-178 is much hotter than the above stars, with $T_{\rm eff} = 9350 \pm 150$ K. For this star we measured over 100 clean, unblended, Fe~{\sc i} and Fe~{\sc ii} lines in the spectral range 500--600 nm. The stellar parameters of $T_{\rm eff}$, $\log g$ and microturbulence were obtained by iteratively adjusting them, using non-linear least squares, in order to find the values which minimized the scatter in the abundance obtained from the Fe lines. This procedure simultaneously attempts to remove any trends in abundance with excitation potential (temperature diagnostic) and equivalent width (microturbulence diagnostic), as well as any differences between the  Fe~{\sc i} and Fe~{\sc ii} lines (surface gravity diagnostic). The parameter uncertainties were obtained from the residual scatter in the optimal solution (see \citealt{2014dapb.book.....N} for further discussion on stellar parameter determination).  

\section{System parameters}
Our process for parametrising the systems combines all our data, photometry and radial-velocity measurements, in one Markov-chain Monte-Carlo (MCMC) analysis, using a code developed in several iterations from that originally described by \citet{2007MNRAS.375..951C}. 

Our standard procedure (see, e.g., \citealt{2019MNRAS.482.1379H}) places a Gaussian ``prior'' on the stellar mass. We derive this using the stellar effective temperature and metallicity, from the spectral analysis, and an estimate of the stellar density, from initial analysis of the transit.  These are used as inputs to the {\sc bagemass} code \citep{2015A&A...575A..36M}, based on  the {\sc garstec} stellar evolution code \citep{2008Ap&SS.316...99W}, which then outputs estimates for the stellar mass and age.  WASP-178 is too hot for the {\sc bagemass} code to be reliable, so we instead adopted a mass prior of 2.04 $\pm$ 0.12 M$_{\odot}$, from expectations of a main-sequence star of its temperature (e.g.~\citealt{2013ApJ...771...40B}), followed by checking that this models the transit to give a self-consistent set of parameters.

In more recent WASP-South papers, following the availability of GAIA DR2 parallaxes \citep{2016A&A...595A...1G,2018A&A...616A...1G}, we also place a prior on the stellar radius. We apply the \citet{2018ApJ...862...61S} correction to the parallax to produce a distance estimate, and then use the Infra-Red Flux Method \citep{1977MNRAS.180..177B} to arrive at the stellar radius.  Before the GAIA DR2, getting the stellar radius wrong was one of the commonest sources of systematic error in transit analyses, and thus a prior on the radius improves the reliability of the solution and can make up for limitations in the transit photometry (see, e.g., \citealt{2019MNRAS.488.3067H}). 

In modelling the RVs we first allowed an eccentric orbit (which is required for WASP-185) but where it was not required (the other three systems) we enforced a circular orbit (as discussed in \citealt{2012MNRAS.422.1988A}, this makes use of the expectation that the time for tidal circularisation of a hot-Jupiter orbit is often shorter than the time in its current orbit).      To fit the transit photometry we adopted limb-darkening coefficients by interpolating from the 4-parameter, non-linear law of \citet{2000A&A...363.1081C}, as appropriate for the star's temperature and metallicity. The WASP passband and the TRAPPIST ``blue block'' filter are wide, non-standard pass bands, for which we approximate by using $R$-band coefficients, which is sufficient for the quality of our photometry.  The MCMC code includes a step where the error bars are inflated such that the fit to each dataset has a $\chi^{2}_{\nu}$ of 1.  This allows for red noise not accounted for in the input errors, thus balancing the different datasets and increasing the output error ranges.  An account of the effects of red noise in typical WASP-planet discovery datasets is given in \citet{2012AJ....143...81S}.

\begin{figure}
\resizebox{\hsize}{!}{\includegraphics{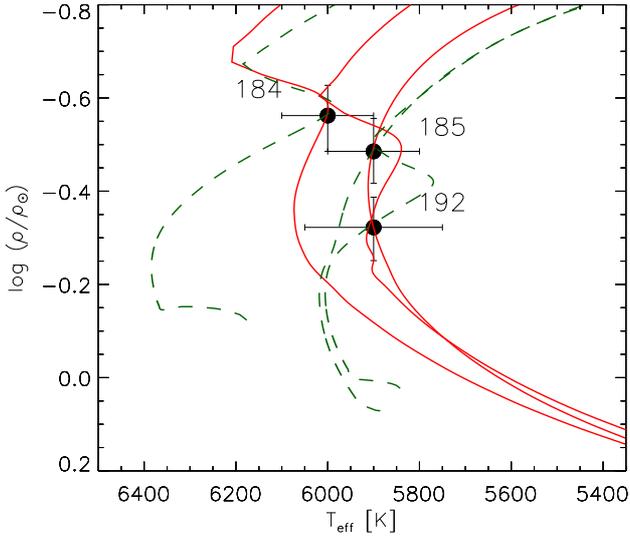}}
\caption{The host star's effective temperature (T$_{\rm eff}$) versus density, each symbol being labelled by the WASP planet number).  We show  best-fit evolution tracks (green dashed lines) and isochrones (red solid lines) for the masses, ages and [Fe/H] values  listed in Table~\ref{ResultsTable}.
\label{trho_plot}
}
\end{figure}

The parameters resulting from the MCMC analysis are listed in the tables for each system. $T_{\rm c}$ is the mid-transit epoch, $P$  is the orbital period, $\Delta F$ is the transit depth that would be observed in the absence of limb-darkening, $T_{14}$ is the duration from first to fourth contact, $b$ the impact parameter, and $K_{\rm 1}$ the stellar reflex velocity.   For WASP-184, WASP-185 and WASP-192 the {\sc bagemass} outputs are tabulated in Table~\ref{ResultsTable} while the best-fit stellar evolution tracks and isochrones are shown in Fig.~\ref{trho_plot} (WASP-178 is too hot for the {\sc bagemass} code to be reliable). 

\begin{table*}
 \caption{Bayesian mass and age estimates for the host stars using {\sc garstec} stellar models assuming $\alpha_{\rm
MLT}=1.78$. Columns 2, 3 and 4 give the maximum-likelihood estimates of the
age, mass, and initial metallicity, respectively. Columns 5 and 6 give the mean and standard deviation of their posterior
distributions. The systematic errors on the mass and age due to
uncertainties in the mixing length and helium abundance are given in Columns 7
to 10.
\label{ResultsTable}}
 \begin{tabular}{@{}lrrrrrrrrrr}
\hline
\hline
Star &
  \multicolumn{1}{c}{$\tau_{\rm iso, b}$ [Gyr]} &
  \multicolumn{1}{c}{$M_{\rm b}$[$M_{\odot}$]} &
  \multicolumn{1}{c}{$\mathrm{[Fe/H]}_\mathrm{i, b}$} &
  \multicolumn{1}{c}{$\langle \tau_{\rm iso} \rangle$ [Gyr]}  &
  \multicolumn{1}{c}{$\langle M_{\star} \rangle$ [$M_{\odot}$]} &
  \multicolumn{1}{c}{$\sigma_{\tau, Y}$}  &
  \multicolumn{1}{c}{$\sigma_{\tau,\alpha}$} &
  \multicolumn{1}{c}{$\sigma_{M, Y}$}  &
  \multicolumn{1}{c}{$\sigma_{M,\alpha}$}  \\
\hline
 \noalign{\smallskip}
WASP-184         &   4.0 &   1.29 & +0.168 & $4.66 \pm  1.15
$&$  1.245 \pm 0.072 $&$ -0.13 $&$  0.40 $&$ -0.036 $&$ -0.005 $ \\
WASP-185         &   7.2 &   1.09 & +0.066 & $6.63 \pm  1.58
$&$  1.116 \pm 0.068 $&$  0.23 $&$  1.87 $&$ -0.048 $&$ -0.069 $ \\
WASP-192         &   5.1 &   1.16 & +0.215 & $5.70 \pm  1.92
$&$  1.137 \pm 0.069 $&$  0.21 $&$  0.98 $&$ -0.043 $&$ -0.016 $ \\
 \noalign{\smallskip}
\hline
\end{tabular}
\end{table*}

\begin{figure}
\hspace*{2mm}\includegraphics[width=8.5cm]{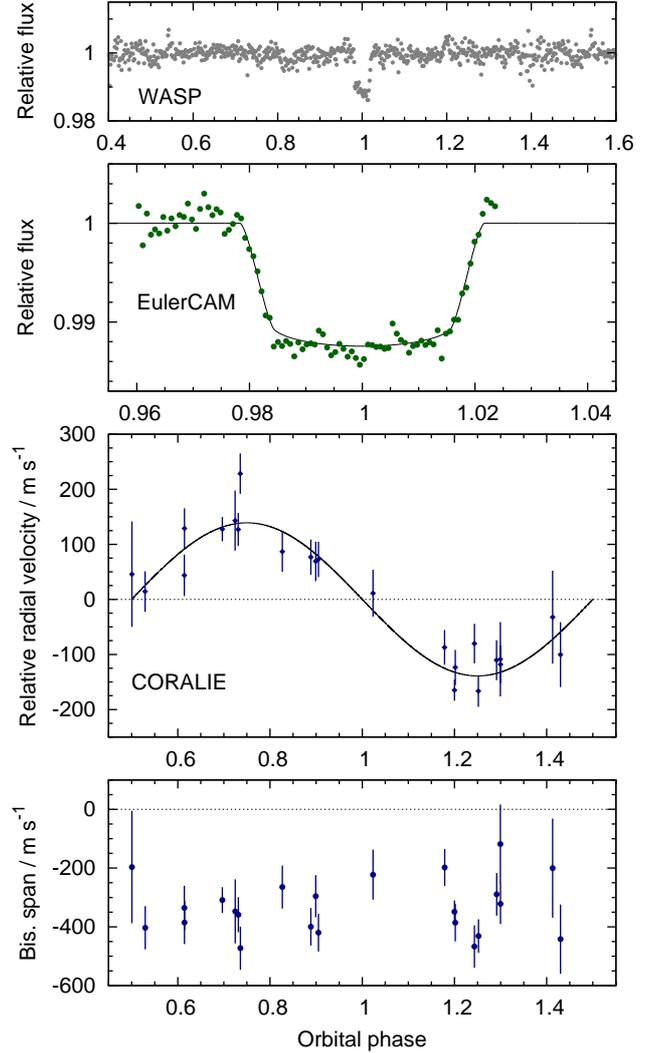}\\ [-2mm]
\caption{WASP-178b discovery data: (Top) The WASP data folded on the 
transit period. (Second panel) The EulerCAM transit lightcurve together with the fitted MCMC model. (Third panel) The CORALIE RV data and fitted model.  (Bottom) The bisector spans of the CORALIE data.}
\end{figure}

\begin{table}
\caption{System parameters for WASP-178.}  
\begin{tabular}{lc}
\multicolumn{2}{l}{1SWASP\,J150904.89-424217.7}\\
\multicolumn{2}{l}{HD\,134004; 2MASS\,15090488--4242178}\\
\multicolumn{2}{l}{GAIA RA\,=\,15$^{\rm h}$09$^{\rm m}$04.89$^{\rm s}$, 
Dec\,=\,--42$^{\circ}$42$^{'}$17.8$^{''}$ (J2000)}\\
\multicolumn{2}{l}{$V$ mag = 9.95; GAIA $G$ = 9.91; $J$ = 9.77}  \\ 
\multicolumn{2}{l}{Rotational modulation: $<$ 1.5 mmag}\\
\multicolumn{2}{l}{GAIA DR2 pm (RA) --10.01\,$\pm$\,0.12 (Dec) --5.65\,$\pm$\,
0.10 mas/yr}\\
\multicolumn{2}{l}{GAIA DR2 parallax: 2.3119 $\pm$ 0.0600 mas}\\
\multicolumn{2}{l}{Distance = 418 $\pm$ 16 pc}\\
\hline
\multicolumn{2}{l}{Stellar parameters from spectroscopic analysis.\rule[-1.5mm]{0mm}{2mm}} \\ \hline 
Spectral type & A1 IV--V \\
$T_{\rm eff}$ (K)  & 9350  $\pm$ 150  \\
$\log g$      & 4.35 $\pm$ 0.15    \\
$v\,\sin i$ (km\,s$^{-1}$)     &   8.2 $\pm$ 0.6     \\
Microturbulence (km\,s$^{-1}$) & 2.9 $\pm$ 0.2 \\
{[Fe/H]}   &  +0.21 $\pm$ 0.16     \\
{[Ca/H]}   &  --0.06 $\pm$ 0.14     \\
{[Sc/H]}   &  --0.35 $\pm$ 0.08     \\
{[Cr/H]}   &  +0.43 $\pm$ 0.10     \\
{[Y/H]}   &  +0.35 $\pm$ 0.10     \\  
{[Ba/H]}   &  +0.96 $\pm$ 0.15     \\
{[Ni/H]}   &  +0.32 $\pm$ 0.12     \\  \hline 
\multicolumn{2}{l}{Parameters from MCMC analysis.\rule[-1.5mm]{0mm}{3mm}} \\
\hline 
$P$ (d) &   3.3448285 $\pm$ 0.0000012 \\
$T_{\rm c}$ (HJD)\,(UTC) & 245\,6927.06839 $\pm$ 0.00047 \\
$T_{\rm 14}$ (d) & 0.1446 $\pm$ 0.0016 \\
$T_{\rm 12}=T_{\rm 34}$ (d) & 0.0197 $\pm$ 0.0016 \\
$\Delta F=R_{\rm P}^{2}$/R$_{*}^{2}$ & 0.01243 $\pm$ 0.00028 \\
$b$ & 0.54 $\pm$ 0.05 \\
$i$ ($^\circ$)  & 85.7 $\pm$ 0.6 \\
$K_{\rm 1}$ (km s$^{-1}$) & 0.139 $\pm$ 0.009 \\
$\gamma$ (km s$^{-1}$)  & $-$23.908 $\pm$ 0.007 \\
$e$ & 0 (adopted) ($<$\,0.08 at 2$\sigma$) \\ 
$a/R_{\rm *}$  & 7.17 $\pm$ 0.21 \\ 
$M_{\rm *}$ (M$_{\rm \odot}$) & 2.07 $\pm$ 0.11 \\
$R_{\rm *}$ (R$_{\rm \odot}$) & 1.67 $\pm$ 0.07 \\
$\log g_{*}$ (cgs) & 4.31 $\pm$ 0.04 \\
$\rho_{\rm *}$ ($\rho_{\rm \odot}$) & 0.44 $\pm$ 0.05\\
$T_{\rm eff}$ (K) & 9360 $\pm$ 150 \\
$M_{\rm P}$ (M$_{\rm Jup}$) & 1.66 $\pm$ 0.12 \\
$R_{\rm P}$ (R$_{\rm Jup}$) & 1.81 $\pm$ 0.09 \\
$\log g_{\rm P}$ (cgs) & 3.07 $\pm$ 0.05 \\
$\rho_{\rm P}$ ($\rho_{\rm J}$) & 0.28 $\pm$ 0.05 \\
$a$ (AU)  & 0.0558  $\pm$ 0.0010 \\
$T_{\rm P, A=0}$ (K) & 2470 $\pm$ 60 \\ [0.5mm] \hline 
\multicolumn{2}{l}{Priors were $M_{\rm *} = 2.04 \pm 0.12\ {\rm M}_{\odot}$ and $R_{\rm *} = 1.81 \pm 0.12\ {\rm R}_{\odot}$}\\
\multicolumn{2}{l}{Errors are 1$\sigma$; Limb-darkening coefficients were:}\\
\multicolumn{2}{l}{{\small $R$ band: a1 = 0.669, a2 = --0.223, a3 = 0.280, a4 = --0.125}}\\ 
\multicolumn{2}{l}{{\small $I$ band: a1 = 0.724, a2 = --0.616, a3 = 0.644, a4 = --0.245}}\\ \hline
\end{tabular} 
\end{table}

\section{WASP-178} 
WASP-178 (= HD 134004) is a bright, $V$ = 9.95, star for which the spectral analysis suggests $T_{\rm eff}$ = 9350 $\pm$ 150 K and an A1 IV--V classification (Table~3; Fig.~2). It appears to be a mild hot Am star, slightly enhanced in Fe ([Fe/H] = +0.21 $\pm$ 0.16)  and slightly depleted in Ca and Sc ([Ca/H] = --0.06 $\pm$ 0.14; [Sc/H] = --0.35 $\pm$ 0.08). Y and Ba are also enhanced by +0.35 $\pm$ 0.10 and +0.96 $\pm$ 0.15, respectively.  Interstellar Na D lines lead to an estimate of $E$($B$--$V$) =   0.06 $\pm$  0.01, which then implies (through the Infra Red Flux Method) a $T_{\rm eff}$ of 9390 $\pm$ 190 K, consistent with that from the spectral analysis.  The projected rotation speed is relatively low at \vsini\ =  8.2  $\pm$ 0.6 \kmps\ (measured assuming zero macroturbulence). We report (Table~3) a stellar mass of 2.07 $\pm$ 0.11 M$_{\odot}$ and a stellar radius of 1.67 $\pm$ 0.07 R$_{\odot}$, which are compatible with a main-sequence, non-evolved status. 

WASP-178 appears to be relatively isolated on the sky, with no nearby stars within 17 arcsec listed in GAIA DR2, and all stars within 30 arcsecs being $>$7 magnitudes fainter. However, WASP-178 is noted in GAIA DR2 as having significant excess noise in the astrometry, amounting to 0.18 mas in 254 astrometric observations.  This could indicate an unresolved and unseen binary companion.  

With a temperature of 9350 $\pm$ 150 K, WASP-178 is the second hottest known host of a hot Jupiter, behind the A0 star KELT-9 \citep{2017Natur.546..514G} at 10\,170 K and ahead of the A2 star MASCARA-2/KELT-20 \citep{2017AJ....154..194L,2018A&A...612A..57T} at 8980 K.  

Despite the high stellar temperature, CORALIE RVs are able to detect the orbital motion.  The planet is in a 3.3-day orbit with a mass of 1.66 $\pm$ 0.12 M$_{\rm Jup}$ and a bloated radius of 1.81 $\pm$ 0.09 R$_{\rm Jup}$. The estimated equilibrium temperature is 2470 $\pm$ 60 K, the hottest of any planet with an orbital period of  $>$\,3 d. Fig.~2 shows the transit photometry and radial-velocity orbit.  We also plot the bisector spans against phase, where the absence of a correlation is a check against transit mimics (e.g. \citealt{2001A&A...379..279Q}).

\begin{figure}
\hspace*{2mm}\includegraphics[width=8.5cm]{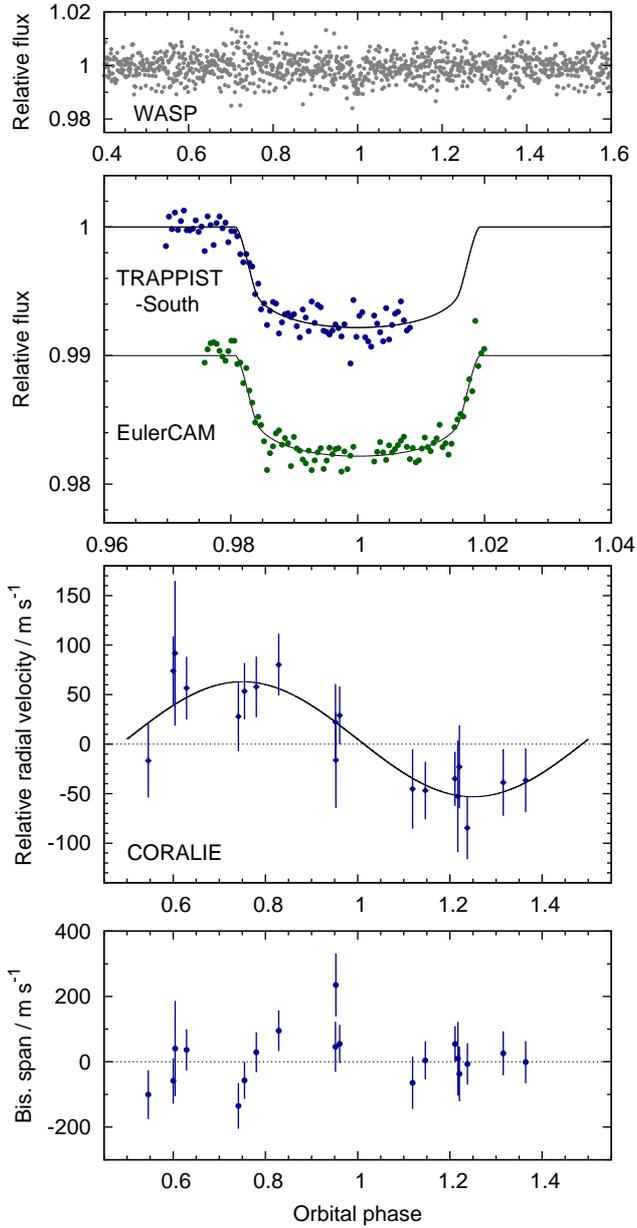}\\ [-2mm]
\caption{WASP-184b discovery data and fitted model, as for Fig.~2. The TRAPPIST, EulerCAM and CORALIE observations are listed in Table~1.}
\end{figure}

\begin{table}
\caption{System parameters for WASP-184.}  
\begin{tabular}{lc}
\multicolumn{2}{l}{1SWASP\,J135804.10--302053.0}\\
\multicolumn{2}{l}{2MASS\,13580408--3020532}\\
\multicolumn{2}{l}{GAIA RA\,=\,13$^{\rm h}$58$^{\rm m}$04.09$^{\rm s}$, 
Dec\,=\,--30$^{\circ}$20$^{'}$53.3$^{''}$ (J2000)}\\
\multicolumn{2}{l}{$V$ mag = 12.9; GAIA $G$ = 12.57; $J$ = 11.6}  \\ 
\multicolumn{2}{l}{Rotational modulation: $<$ 2 mmag}\\
\multicolumn{2}{l}{GAIA DR2 pm (RA) --4.36\,$\pm$\,0.06 (Dec) --5.09\,$\pm$\,
0.06 mas/yr}\\
\multicolumn{2}{l}{GAIA DR2 parallax: 1.480 $\pm$ 0.037 mas}\\
\multicolumn{2}{l}{Distance = 640 $\pm$ 28 pc}\\
\hline
\multicolumn{2}{l}{Stellar parameters from spectroscopic analysis.\rule[-1.5mm]{0mm}{2mm}} \\ \hline 
Spectral type & G0 \\
$T_{\rm eff}$ (K)  & 6000  $\pm$ 100  \\
$\log g$      & 4.0 $\pm$ 0.2    \\
$v\,\sin i$ (km\,s$^{-1}$)     &   4.5 $\pm$ 1.1     \\
{[Fe/H]}   &  +0.12 $\pm$ 0.08     \\
log A(Li)  &  2.04 $\pm$ 0.08    \\  \hline 
\multicolumn{2}{l}{Parameters from MCMC analysis.\rule[-1.5mm]{0mm}{3mm}} \\
\hline 
$P$ (d) &   5.18170 $\pm$ 0.00001 \\
$T_{\rm c}$ (HJD)\,(UTC) & 245\,7630.008 $\pm$ 0.001 \\
$T_{\rm 14}$ (d) & 0.1990 $\pm$ 0.0027 \\
$T_{\rm 12}=T_{\rm 34}$ (d) & 0.0187 $\pm$ 0.0024 \\
$\Delta F=R_{\rm P}^{2}$/R$_{*}^{2}$ & 0.0069 $\pm$ 0.0003 \\
$b$ & 0.44 $\pm$ 0.14 \\
$i$ ($^\circ$)  & 86.9 $\pm$ 1.1 \\
$K_{\rm 1}$ (km s$^{-1}$) & 0.058 $\pm$ 0.010 \\
$\gamma$ (km s$^{-1}$)  & 8.366 $\pm$ 0.008 \\
$e$ & 0 (adopted) ($<$\,0.25 at 2$\sigma$) \\ 
$a/R_{\rm *}$  & 8.19 $\pm$ 0.42 \\ 
$M_{\rm *}$ (M$_{\rm \odot}$) & 1.23 $\pm$ 0.07 \\
$R_{\rm *}$ (R$_{\rm \odot}$) & 1.65 $\pm$ 0.09 \\
$\log g_{*}$ (cgs) & 4.09 $\pm$ 0.05 \\
$\rho_{\rm *}$ ($\rho_{\rm \odot}$) & 0.27 $\pm$ 0.05\\
$T_{\rm eff}$ (K) & 6000 $\pm$ 100 \\
$M_{\rm P}$ (M$_{\rm Jup}$) & 0.57 $\pm$ 0.10 \\
$R_{\rm P}$ (R$_{\rm Jup}$) & 1.33 $\pm$ 0.09 \\
$\log g_{\rm P}$ (cgs) & 2.87 $\pm$ 0.10 \\
$\rho_{\rm P}$ ($\rho_{\rm J}$) & 0.24 $\pm$ 0.07 \\
$a$ (AU)  & 0.0627  $\pm$ 0.0012 \\
$T_{\rm P, A=0}$ (K) & 1480 $\pm$ 50 \\ [0.5mm] \hline
\multicolumn{2}{l}{Priors were $M_{\rm *} = 1.25 \pm 0.07\ {\rm M}_{\odot}$ and $R_{\rm *} = 1.59 \pm 0.10\ {\rm R}_{\odot}$}\\ 
\multicolumn{2}{l}{Errors are 1$\sigma$; Limb-darkening coefficients were:}\\
\multicolumn{2}{l}{{\small $R$ band: a1 = 0.578, a2 = 0.022, a3 = 0.359, a4 = --0.230}}\\ \hline
\end{tabular} 
\end{table}

\section{WASP-184}
WASP-184 is a $V$ = 12.9, G0 star with a metallicity of [Fe/H] = +0.12 $\pm$ 0.08 and a distance of 640 $\pm$ 28 pc (Table~4; Fig.~3).   WASP-184 is relatively isolated with no stars recorded in GAIA DR2 within 10 arcsecs, and only 2 stars ($>$6 magnitudes fainter) within 30 arcsecs. There is no excess astrometric noise recorded in DR2.   The mass and radius of WASP-184 (1.23 $\pm$ 0.07 M$_{\odot}$; 1.65 $\pm$ 0.09 R$_{\odot}$) imply that it is evolving off the main sequence. Using the {\sc bagemass} code we compute an age of 4.7 $\pm$ 1.1 Gyr. Lithium depletion to the measured value of log A(Li)= 2.04 $\pm$ 0.08 could take $\sim$\,5 Gyr according to Table~3 of \citet{2005A&A...442..615S}, which is consistent with the {\sc bagemass} age. 

The system is reasonably well parametrised by a partial transit from  TRAPPIST-South, a nearly-full transit from EulerCAM, and 19 RVs from CORALIE.  The planet is in a 5.18-d orbit and is a moderately bloated, lower-mass hot Jupiter (0.57 $\pm$ 0.07 M$_{\rm Jup}$; 1.33 $\pm$ 0.09 R$_{\rm Jup}$).  

\begin{figure}
\hspace*{2mm}\includegraphics[width=8.5cm]{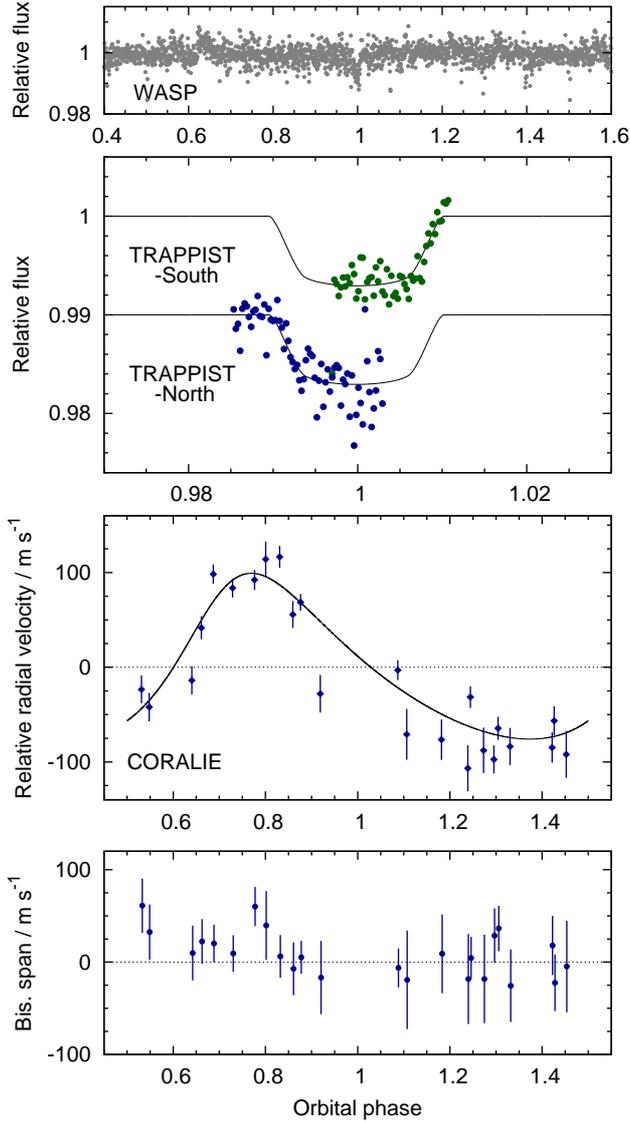}\\ [-2mm]
\caption{WASP-185b discovery data and fitted model, as for Figs.~2 \&\ 3.}
\end{figure}

\begin{table}
\caption{System parameters for WASP-185.}  
\begin{tabular}{lc}
\multicolumn{2}{l}{1SWASP\,J141614.30--193232.1}\\
\multicolumn{2}{l}{2MASS\,14161431--1932321}\\
\multicolumn{2}{l}{GAIA RA\,=\,14$^{\rm h}$16$^{\rm m}$14.31$^{\rm s}$, 
Dec\,=\,--19$^{\circ}$32$^{'}$32.2$^{''}$ (J2000)}\\
\multicolumn{2}{l}{$V$ mag = 11.1; GAIA $G$ = 10.89; $J$ = 9.87}  \\ 
\multicolumn{2}{l}{Rotational modulation: $<$ 1 mmag}\\
\multicolumn{2}{l}{GAIA DR2 pm (RA) --13.40\,$\pm$\,0.08 (Dec) --6.06\,$\pm$\,
0.07 mas/yr}\\
\multicolumn{2}{l}{GAIA DR2 parallax: 3.552 $\pm$ 0.043 mas}\\
\multicolumn{2}{l}{Distance = 275 $\pm$ 6 pc}\\
\hline
\multicolumn{2}{l}{Stellar parameters from spectroscopic analysis.\rule[-1.5mm]{0mm}{2mm}} \\ \hline 
Spectral type & G0 \\
$T_{\rm eff}$ (K)  & 5900  $\pm$ 100  \\
$\log g$      & 4.0 $\pm$ 0.2    \\
$v\,\sin i$ (km\,s$^{-1}$)     &   2.8 $\pm$ 0.9     \\
{[Fe/H]}   &  --0.02 $\pm$ 0.06     \\
log A(Li)  &  2.37 $\pm$ 0.09    \\ \hline 
\multicolumn{2}{l}{Parameters from MCMC analysis.\rule[-1.5mm]{0mm}{3mm}} \\
\hline 
$P$ (d) &   9.38755 $\pm$ 0.00002 \\
$T_{\rm c}$ (HJD)\,(UTC) & 245\,6935.982 $\pm$ 0.002 \\
$T_{\rm 14}$ (d) & 0.192 $\pm$ 0.006 \\
$T_{\rm 12}=T_{\rm 34}$ (d) & 0.040 $\pm$ 0.006 \\
$\Delta F=R_{\rm P}^{2}$/R$_{*}^{2}$ & 0.0073 $\pm$ 0.0005 \\
$b$ & 0.81 $\pm$ 0.03 \\
$i$ ($^\circ$)  & 86.8 $\pm$ 0.3 \\
$K_{\rm 1}$ (km s$^{-1}$) & 0.088 $\pm$ 0.004 \\
$\gamma$ (km s$^{-1}$)  & 23.874  $\pm$ 0.003 \\
$e$ & 0.24 $\pm$ 0.04  \\ 
$\omega$ (deg) & --42 $\pm$ 7 \\ 
$a/R_{\rm *}$  & 12.9 $\pm$ 0.7 \\ 
$M_{\rm *}$ (M$_{\rm \odot}$) & 1.12 $\pm$ 0.06 \\
$R_{\rm *}$ (R$_{\rm \odot}$) & 1.50 $\pm$ 0.08 \\
$\log g_{*}$ (cgs) & 4.13 $\pm$ 0.05 \\
$\rho_{\rm *}$ ($\rho_{\rm \odot}$) & 0.33 $\pm$ 0.06\\
$T_{\rm eff}$ (K) & 5900 $\pm$ 100 \\
$M_{\rm P}$ (M$_{\rm Jup}$) & 0.98 $\pm$ 0.06 \\
$R_{\rm P}$ (R$_{\rm Jup}$) & 1.25 $\pm$ 0.08 \\
$\log g_{\rm P}$ (cgs) & 3.15 $\pm$ 0.07 \\
$\rho_{\rm P}$ ($\rho_{\rm J}$) & 0.50 $\pm$ 0.12 \\
$a$ (AU)  & 0.0904  $\pm$ 0.0017 \\
$T_{\rm P, A=0}$ (K) & 1160 $\pm$ 35 \\ [0.5mm] \hline 
\multicolumn{2}{l}{Priors were $M_{\rm *} = 1.11 \pm 0.06\ {\rm M}_{\odot}$ and $R_{\rm *} = 1.58 \pm 0.09\ {\rm R}_{\odot}$}\\
\multicolumn{2}{l}{Errors are 1$\sigma$; Limb-darkening coefficients were:}\\
\multicolumn{2}{l}{{\small $R$ band: a1 = 0.568, a2 = --0.009, a3 = 0.443, a4 = --0.271}}\\ 
\multicolumn{2}{l}{{\small $z$ band: a1 = 0.651, a2 = --0.334, a3 = 0.621, a4 = --0.320}}\\ \hline
\end{tabular} 
\end{table}

\section{WASP-185}
WASP-185 is a $V$ = 11.1, G0 star with a solar metallicity ([Fe/H] = --0.02 $\pm$ 0.06) at a distance of 275 $\pm$ 6 pc (Table~5; Fig.~4).    It has an apparent companion star 4.6 arcsecs away and 4.4 mag fainter in GAIA $G$ (too faint for GAIA to report its proper motion, so we don't know whether the two are physically associated; at the distance of WASP-185 the separation would correspond to 1200 AU). Otherwise WASP-185 is relatively isolated (with 3 other stars, $>$9 magnitudes fainter, between 20 and 30 arcsecs away). There is no DR2 excess astrometric noise reported for WASP-185.     The companion star is sufficiently distant that it will not affect the CORALIE RVs, however it is included in the extraction aperture for the TRAPPIST photometry. We therefore applied a correction of 1.8\%\ to the transit photometry, though in practice this amount is much less than the uncertainties. 

The mass and radius of WASP-185 (1.12 $\pm$ 0.06 M$_{\odot}$; 1.50 $\pm$ 0.08 R$_{\odot}$) indicate an evolved star, and the {\sc bagemass} code suggests an age of 6.6 $\pm$ 1.6 Gyr.  Lithium depletion to the measured value of log A(Li) = 2.37 $\pm$ 0.09 could take $\sim$\,2 Gyr, but this abundance of lithium is found in NGC 188 which is $\sim$\,8 Gyr old according to Table~3 of \citet{2005A&A...442..615S}. Thus the lithium is consistent with the {\sc bagemass} age. 

We have only limited photometry of the transit, one ingress and one egress, both obtained in deteriorating observing conditions, and so the transit parameterisation depends substantially on the stellar radius deduced from the GAIA DR2 distance. Our 24 CORALIE RVs trace out an eccentric orbit, though there is clearly additional scatter of unknown origin.  This could be magnetic activity of the host star, though no rotational modulation is seen in the WASP data to a limit of 1 mmag (WASP-166 is an example of a system showing RV variation owing to magnetic activity, but no rotational modulation; \citealt{2019MNRAS.488.3067H}). 

The planet's orbit has a relatively long period for a hot Jupiter, at 9.39 d, and has an eccentricity of $e$ = 0.24 $\pm$ 0.04. The impact factor is relatively high at $b$ = 0.81 $\pm$ 0.03.   The planet's mass and radius (0.98 $\pm$ 0.06 M$_{\rm Jup}$; 1.25 $\pm$ 0.08 R$_{\rm Jup}$) are typical for hot Jupiters.

\begin{figure}
\hspace*{2mm}\includegraphics[width=8.5cm]{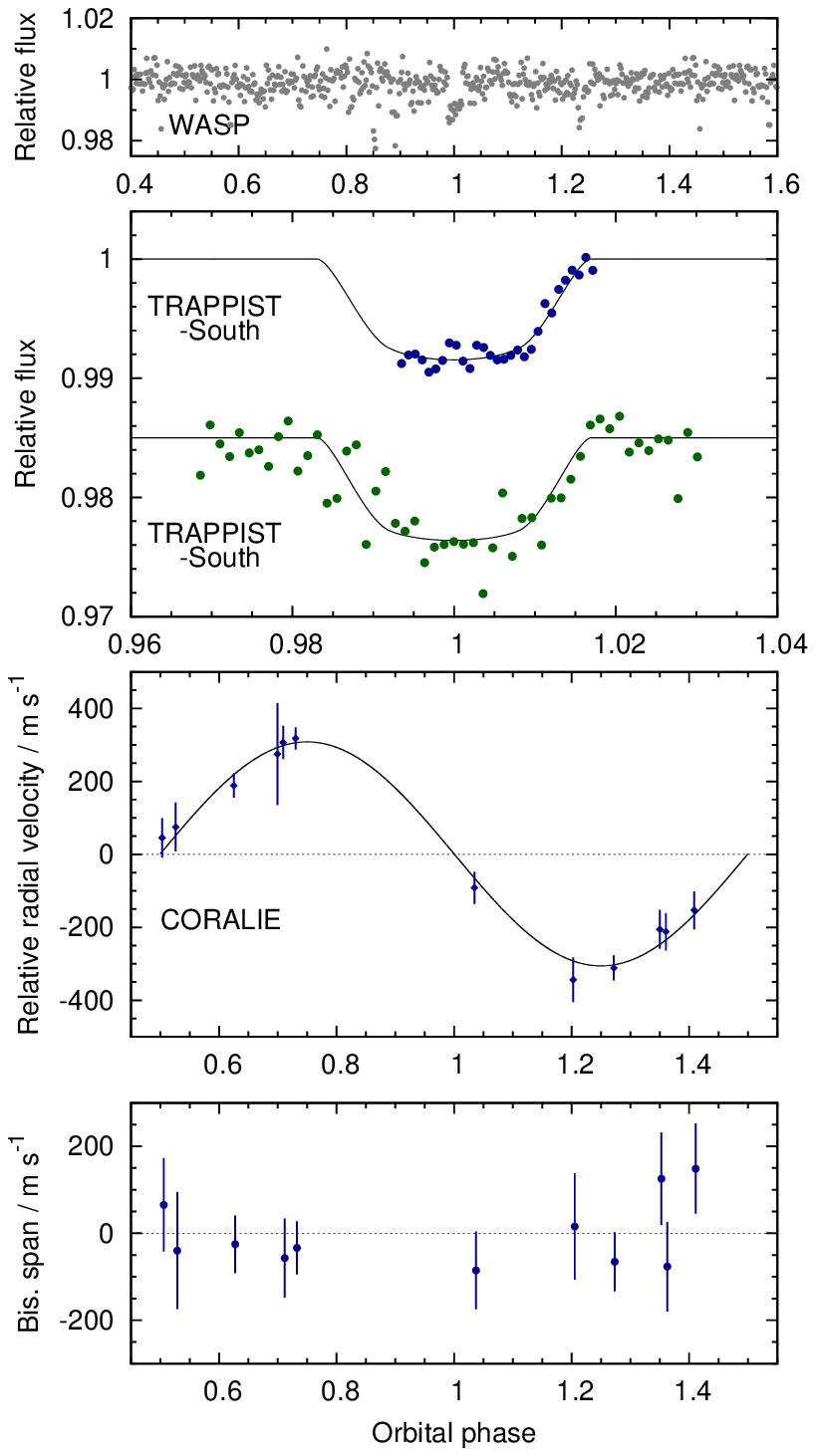}\\ [-2mm]
\caption{WASP-192b discovery data and fitted model, as for Figs.~2 \&\ 3.}
\end{figure}

\begin{table}
\caption{System parameters for WASP-192.}
\begin{tabular}{lc}
\multicolumn{2}{l}{1SWASP\,J145438.06--384439.6}\\
\multicolumn{2}{l}{2MASS\,14543809--3844403}\\
\multicolumn{2}{l}{GAIA RA\,=\,14$^{\rm h}$54$^{\rm m}$38.09$^{\rm s}$, 
Dec\,=\,--38$^{\circ}$44$^{'}$40.3$^{''}$ (J2000)}\\
\multicolumn{2}{l}{$V$ mag = 12.3; GAIA $G$ = 12.53; $J$ = 11.5}  \\ 
\multicolumn{2}{l}{Rotational modulation: $<$ 2  mmag}\\
\multicolumn{2}{l}{GAIA DR2 pm (RA) 0.72\,$\pm$\,0.07 (Dec) --1.53\,$\pm$\,
0.06 mas/yr}\\
\multicolumn{2}{l}{GAIA DR2 parallax: 1.939 $\pm$ 0.062 mas}\\
\multicolumn{2}{l}{Distance = 495 $\pm$ 22 pc}\\
\hline
\multicolumn{2}{l}{Stellar parameters from spectroscopic analysis.\rule[-1.5mm]{0mm}{2mm}} \\ \hline 
Spectral type & G0 \\
$T_{\rm eff}$ (K)  & 5900  $\pm$ 150  \\
$\log g$      & 4.5 $\pm$ 0.2    \\
$v\,\sin i$ (km\,s$^{-1}$)     &   3.1 $\pm$ 1.1     \\
{[Fe/H]}   &  +0.14 $\pm$ 0.08     \\
log A(Li)  &  2.11 $\pm$ 0.13    \\ \hline 
\multicolumn{2}{l}{Parameters from MCMC analysis.\rule[-1.5mm]{0mm}{3mm}} \\
\hline 
$P$ (d) & 2.8786765 $\pm$ 0.0000028 \\
$T_{\rm c}$ (HJD)\,(UTC) & 245\,7271.3331 $\pm$ 0.0017 \\
$T_{\rm 14}$ (d) & 0.0964 $\pm$ 0.0040 \\
$T_{\rm 12}=T_{\rm 34}$ (d) & 0.026 $\pm$ 0.004 \\
$\Delta F=R_{\rm P}^{2}$/R$_{*}^{2}$ & 0.00926 $\pm$ 0.00061 \\
$b$ & 0.84 $\pm$ 0.03 \\
$i$ ($^\circ$)  & 82.7 $\pm$ 0.6 \\
$K_{\rm 1}$ (km s$^{-1}$) & 0.307 $\pm$ 0.017 \\
$\gamma$ (km s$^{-1}$)  & 15.896 $\pm$ 0.013 \\
$e$ & 0 (adopted) ($<$\,0.25 at 2$\sigma$) \\ 
$a/R_{\rm *}$  & 6.65 $\pm$ 0.34 \\ 
$M_{\rm *}$ (M$_{\rm \odot}$) & 1.09 $\pm$ 0.06 \\
$R_{\rm *}$ (R$_{\rm \odot}$) & 1.32 $\pm$ 0.07 \\
$\log g_{*}$ (cgs) & 4.236 $\pm$ 0.051 \\
$\rho_{\rm *}$ ($\rho_{\rm \odot}$) & 0.476 $\pm$ 0.080\\
$T_{\rm eff}$ (K) & 5910 $\pm$ 145 \\
$M_{\rm P}$ (M$_{\rm Jup}$) & 2.30 $\pm$ 0.16 \\
$R_{\rm P}$ (R$_{\rm Jup}$) & 1.23 $\pm$ 0.08 \\
$\log g_{\rm P}$ (cgs) & 3.54 $\pm$ 0.07 \\
$\rho_{\rm P}$ ($\rho_{\rm J}$) & 1.22 $\pm$ 0.31 \\
$a$ (AU)  & 0.0408  $\pm$ 0.0008 \\
$T_{\rm P, A=0}$ (K) & 1620 $\pm$ 60 \\ [0.5mm] \hline 
\multicolumn{2}{l}{Priors were $M_{\rm *} = 1.09 \pm 0.06\ {\rm M}_{\odot}$ and $R_{\rm *} = 1.34 \pm 0.08\ {\rm R}_{\odot}$}\\
\multicolumn{2}{l}{Errors are 1$\sigma$; Limb-darkening coefficients were:}\\
\multicolumn{2}{l}{{\small $R$ band: a1 = 0.621, a2 = --0.179, a3 = 0.655, a4 = --0.356}}\\ 
\multicolumn{2}{l}{{\small $I$ band: a1 =   0.697, a2 = --0.435, a3 = 0.801, a4 = --0.394}}\\ \hline
\end{tabular} 
\end{table}

\begin{table*}
\caption{The hottest of all known ultra-hot Jupiters}
\begin{tabular}{lccrcccl}
Name & Eq.~Temp & Host & Host  & Period  & Radius  & Mass  &  Discovery \\ 
& (K) & &  \mbox{\ \ $V$\ \ } & (d) & (Jup) & (Jup) \\
[0.5mm] \hline
KELT-9b & 4050 & A0 & 7.6 & 1.48 & 1.89 & 2.9 & \citet{2017Natur.546..514G}\\
WASP-33b & 2780 & A5 & 8.3 & 1.22 & 1.60 & 2.1 & \citet{2010MNRAS.407..507C}\\
Kepler-13b & 2750 & A2 & 10.0 & 1.76 & 1.41 & $\sim$9 & \citet{2011AJ....142..195S} \\
WASP-189b & 2640 & A6 & 6.6 & 2.72 & 1.40 & 1.9 & \citet{2018arXiv180904897A}\\
WASP-12b & 2590 & G0 & 11.7 & 1.09 & 1.90 & 1.5 & \citet{2009ApJ...693.1920H}\\
MASCARA-1b & 2570 & A8 & 8.3 & 2.15 & 1.50 & 3.7 & {\citet{2017A&A...606A..73T}}\\
HAT-P-70b & 2560 & & 9.5  & 2.74 & 1.87 & &  \citet{2019arXiv190600462Z}\\
WASP-103b & 2510 & F8 & 12.0 & 0.92  & 1.53 & 1.5 & {\citet{2014A&A...562L...3G}}\\ 
WASP-178b & 2470 & A1 & 9.9 & 3.34 & 1.81 & 1.7 & This work \\
WASP-78b & 2470 & F8 & 12.0 & 2.17 & 2.06 & 0.9 & {\citet{2012A&A...547A..61S}}\\ 
KELT-16b & 2450 &F7  & 11.9 & 0.97 & 1.42 & 2.7 & \citet{2017AJ....153...97O} \\ 
WASP-18b & 2410 & F9 & 9.3 & 0.94 & 1.20 & 10.5 & \citet{2009Natur.460.1098H} \\ 
WASP-121b & 2360 & F6 & 10.4 & 1.27 & 1.87 & 1.2 & \citet{2016MNRAS.458.4025D}\\
WASP-167b/KELT-13b & 2330 & F1 & 10.5 & 2.02 & 1.51 & & \citet{2017MNRAS.471.2743T}\\
WASP-87Ab & 2320 & F5 & 10.7 & 1.68 & 1.39 & 2.2 & \citet{2014arXiv1410.3449A}\\
\end{tabular} 
\end{table*}

\section{WASP-192}
 WASP-192 is  a $V$ = 12.3, G0 star with metallicity [Fe/H] = +0.14 $\pm$ 0.08 at a distance of 495 $\pm$ 22 pc (Table~6; Fig.~5). It is isolated in the sky, with no stars, less than 6 magnitudes fainter, within 30 arcsecs according to GAIA DR2. There is no excess astrometric noise reported in DR2.  The mass and radius (1.09 $\pm$ 0.06 M$_{\odot}$; 1.32 $\pm$ 0.07 R$_{\odot}$) indicate a moderately evolved star, and the {\sc bagemass} code produces an age of 5.7 $\pm$ 1.9 Gyr.  Lithium depletion to the measured log A(Li) =  2.11 $\pm$ 0.13 could take $\sim$\,5 Gyr according to Table~3 of \citet{2005A&A...442..615S}, which is consistent with the {\sc bagemass} age. 

The planet WASP-192b has a typical hot-Jupiter orbit of $P$ = 2.88 d with a relatively high impact parameter of $b$ = 0.84 $\pm$ 0.03.  We have TRAPPIST photometry of one partial transit and one full transit, though that was in poorer observing conditions. The planet is more massive than average for a hot Jupiter at 2.30 $\pm$ 0.16 M$_{\rm Jup}$, such that 12 CORALIE RVs show a well-defined orbital motion. The radius of 1.23 $\pm$ 0.08 R$_{\rm Jup}$ is typical of hot Jupiters that have masses in the range 2--3 M$_{\rm Jup}$.

\section{Discussion}
Recent papers have outlined a class of ``ultra-hot Jupiters'', defined by \citet{2018A&A...617A.110P} as Jupiters with day-side temperatures greater than 2200 K.   Atmospheric characterisation of UHJs such as WASP-18b, WASP-103b and WASP-121b (e.g.~\citealt{2018AJ....156...17K,2019A&A...625A.136A}) has revealed systematically different behaviour from cooler planets.  Whereas cooler planets can show strong water features (e.g.~WASP-107b; \citealt{2018ApJ...858L...6K}) water is thought to disassociate on the day-sides of UHJs, such that no water features are seen. The disassociated ions then drift to the night side, where they recombine. The molecule CO, however, has a stronger molecular bond, and is still present on the day sides of UHJs, where it can produce an emission feature (e.g.~\citealt{2018A&A...617A.110P}). 

In Table~7 we list the hottest of all the known UHJs, those with a calculated equilibrium temperature above 2300 K (the UHJ definition of {\it day-side\/} temperature $>$\,2200 K includes many more objects than we list). We use equilibrium temperature,  taking the data from  TEPcat\footnote{https://www.astro.keele.ac.uk/jkt/tepcat/ \citep{2011MNRAS.417.2166S},  since it can be calculated uniformly for all the known systems.} WASP-178b now joins this group. Transiting a $V$ = 9.95 star, it is among the best UHJ targets visible from the Southern Hemisphere, along with WASP-18b, WASP-103b, WASP-121b and WASP-189b. 

A correlation between high irradiation of hot Jupiters and bloated radii is now well established (e.g.~\citealt{2016AJ....152..182H,2016arXiv160700322B,2018A&A...616A..76S}).  WASP-178b is at the upper end of such a relationship, as illustrated for known planets in Fig.~6.  Also apparent in Table~7 is a tendency for the hottest HJs to be more massive than typical.  The median mass of a transiting hot Jupiter is $\sim$\,0.9 M$_{\rm Jup}$, whereas the median of those in Table~7 is 2.2 M$_{\rm Jup}$.  This presumably reflects the destruction of irradiated gas giants by photo-evaporation (e.g.~\citealt{2018MNRAS.479.5012O}), such that lower-mass UHJs would have short lifetimes.  Indeed, lower-mass UHJs such as WASP-12b (1.5 M$_{\rm Jup}$) are seen to be losing mass (e.g.~\citealt{2013ApJ...766L..20F}). With a moderate mass of 1.7 M$_{\rm Jup}$, WASP-178b is thus also a candidate for photo-evaporation.

\begin{figure}
\hspace*{2mm}\includegraphics[width=8.5cm]{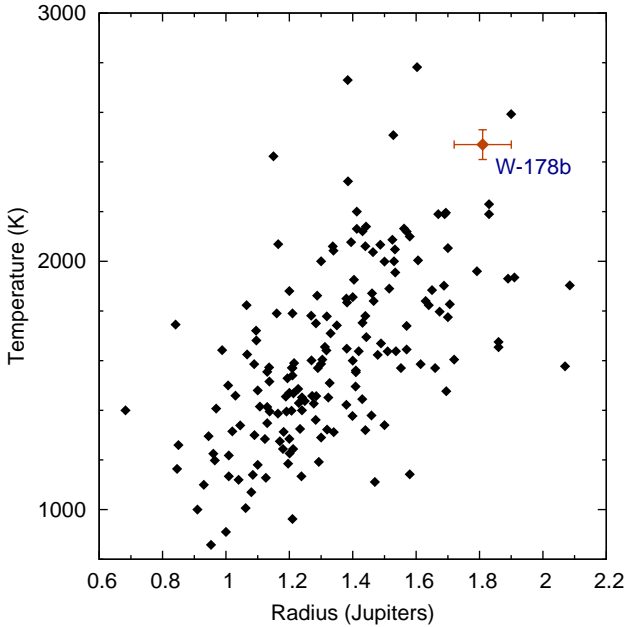}\\ [-2mm]
\caption{Radii and calculated temperatures of transiting hot Jupiters, showing the location of WASP-178b in red.  KELT-9b is above the plot at 4600 K.   The data are from http://exoplanet.eu.  We caution about selection effects in such a plot since non-bloated planets would have shallower transits against larger, hotter stars, so would be harder to detect.}
\end{figure}

We turn now to WASP-185b, which is notable for its eccentric orbit of $e = 0.24$. The tidal circularisation timescale increases markedly with orbital period, and so eccentric orbits are more likely for longer periods such as WASP-185b's 9.39 d (see Fig.~7). Using eqn 3 of \citet{2006ApJ...649.1004A} we can estimate the circularisation timescale of WASP-185b as $\sim$\,2 Gyr, though this depends on assuming $Q_{P}$\,$\sim$\,10$^{5}$, which is uncertain.  There is a tendency, however, for hot Jupiters with eccentric orbits to be either more massive (e.g.\ WASP-8b at 2.2 M$_{\rm Jup}$; \citealt{2010A&A...517L...1Q}, WASP-162b at 5.2 M$_{\rm Jup}$; \citealt{2019MNRAS.482.1379H}, and HAT-P-34b at 3.3 M$_{\rm Jup}$; \citealt{2012AJ....144...19B}) or to have indications of additional bodies in the system that might be perturbing the hot Jupiter (e.g.\ HAT-P-31b,c; \citealt{2011AJ....142...95K} and HAT-P-17b,c; \citealt{2012ApJ...749..134H}).  Given that WASP-185b is only 1 M$_{\rm Jup}$, and so should circularise more rapidly, and given the relatively long 6.6 $\pm$ 1.6 Gyr age of the host star, it may be that WASP-185b has arrived in its current orbit more recently, or that it is being perturbed by an outer companion (e.g.~\citealt{2016ApJ...829..132P}), possibly the putative companion at 1200 AU.   It would thus be worthwhile to obtain Rossiter--McLaughlin observations of WASP-185b to discern whether the planet's orbit is aligned or mis-aligned with the stellar rotation. 

\begin{figure}
\hspace*{0mm}\includegraphics[width=9cm]{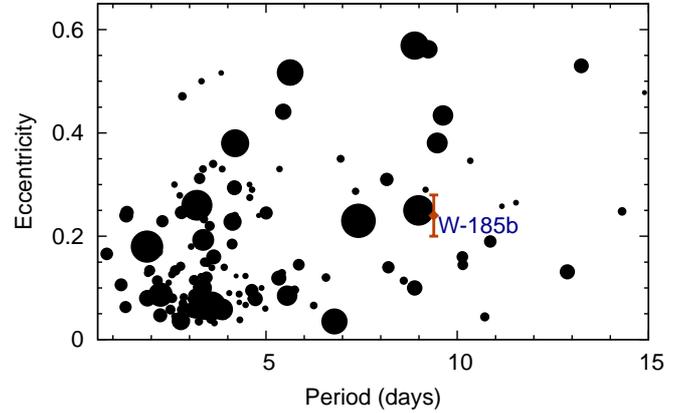}\\ [-2mm]
\caption{Orbital eccentricity versus orbital period for hot Jupiters, with WASP-185b's location in red. Only systems with $e > 0.03$ are shown. The symbol area scales with the planet mass. Data are from http://exoplanet.eu.}
\end{figure}

\section*{Acknowledgements}
WASP-South was hosted by the South African Astronomical Observatory
and we are grateful for their support and assistance. Funding for WASP
came from consortium universities and from the UK's Science and
Technology Facilities Council. The Euler Swiss telescope is supported
by the Swiss National Science Foundation. The research leading to
these results has received funding from the ARC grant for Concerted
Research Actions, financed by the Wallonia-Brussels
Federation. TRAPPIST-South is funded by the Belgian Fund for
Scientific Research (Fond National de la Recherche Scientifique, FNRS)
under the grant FRFC 2.5.594.09.F, with the participation of the Swiss
National Science Fundation (SNF). MG and EJ are F.R.S.-FNRS Senior
Research Associates.

\bibliographystyle{mnras}
\bibliography{biblio}

\newpage
\vspace*{-2cm}

\begin{table}
\renewcommand\thetable{A1}
\caption{Radial velocities.\protect\rule[-1.5mm]{0mm}{2mm}} 
\begin{tabular}{cccr} 
\hline 
BJD\,--\,2400\,000 & RV & $\sigma_{\rm RV}$ & Bisector \\
(UTC)  & (km s$^{-1}$) & (km s$^{-1}$) & (km s$^{-1}$)\\ [0.5mm] \hline
\multicolumn{4}{l}{{\bf WASP-178:}}\\  
57850.91018 &   $-$24.0725 &    0.0192 &        $-$0.3486 \\
57893.80235 &   $-$23.8966 &    0.0426 &        $-$0.2225 \\
57894.56659 &   $-$24.0744 &    0.0285 &        $-$0.4313 \\
57904.73294 &   $-$24.0182 &    0.0364 &        $-$0.2893 \\
57934.67540 &   $-$23.9881 &    0.0360 &        $-$0.4670 \\
57949.66427 &   $-$23.7647 &    0.0545 &        $-$0.3473 \\
57951.58735 &   $-$24.0164 &    0.0675 &        $-$0.1183 \\
57952.64066 &   $-$23.8640 &    0.0377 &        $-$0.3355 \\
57954.60536 &   $-$24.0314 &    0.0319 &        $-$0.3859 \\
57955.60534 &   $-$23.8620 &    0.0957 &        $-$0.1964 \\
57958.65602 &   $-$23.9400 &    0.0843 &        $-$0.2004 \\
57959.60533 &   $-$23.7802 &    0.0220 &        $-$0.3087 \\
57974.59726 &   $-$23.9948 &    0.0316 &        $-$0.1979 \\
58002.52881 &   $-$23.8935 &    0.0367 &        $-$0.4032 \\
58018.48523 &   $-$24.0258 &    0.0345 &        $-$0.3217 \\
58020.49026 &   $-$23.8387 &    0.0359 &        $-$0.2958 \\
58030.48938 &   $-$23.8311 &    0.0320 &        $-$0.3998 \\
58203.89388 &   $-$23.7806 &    0.0299 &        $-$0.3586 \\
58207.82049 &   $-$23.8350 &    0.0322 &        $-$0.4199 \\
58247.69603 &   $-$23.8210 &    0.0366 &        $-$0.2646 \\
58276.47209 &   $-$24.0082 &    0.0590 &        $-$0.4419 \\
58277.49486 &   $-$23.6794 &    0.0367 &        $-$0.4725 \\
58320.57235 &   $-$23.7790 &    0.0367 &        $-$0.3855 \\ [0.5mm]
\hline
%
%
\multicolumn{4}{l}{{\bf WASP-184:}}\\  
57190.68828 &   8.3083 & 0.0564 &    0.0095 \\
57618.50533 &   8.4188 & 0.0306 &    0.0288 \\
57817.78113 &   8.2765 & 0.0317 & $-$0.0066 \\
57905.73109 &   8.3261 & 0.0273 &    0.0545 \\
57924.47781 &   8.4413 & 0.0312 &    0.0950 \\
57933.67645 &   8.4528 & 0.0730 &    0.0404 \\
57954.53219 &   8.4176 & 0.0317 &    0.0363 \\
57959.56409 &   8.4350 & 0.0347 & $-$0.0586 \\
58170.79185 &   8.3245 & 0.0323 & $-$0.0012 \\
58171.73350 &   8.3443 & 0.0372 & $-$0.1005 \\
58172.74625 &   8.3889 & 0.0350 & $-$0.1351 \\
58173.88416 &   8.3900 & 0.0292 &    0.0547 \\
58174.70319 &   8.3157 & 0.0400 & $-$0.0645 \\
58175.72199 &   8.3224 & 0.0337 &    0.0258 \\
58247.77332 &   8.3381 & 0.0419 & $-$0.0371 \\
58277.47072 &   8.3834 & 0.0384 &    0.0459 \\
58307.53963 &   8.4145 & 0.0284 & $-$0.0570 \\
58308.56619 &   8.3448 & 0.0482 &    0.2351 \\
58309.57037 &   8.3142 & 0.0291 &    0.0041 \\
58593.66628 &   8.0600 & 0.0625 &    0.0982 \\ [0.5mm]
\hline
\multicolumn{4}{l}{Bisector errors are twice RV errors} 
\end{tabular} 
\end{table} 

\begin{table}
\begin{tabular}{cccr} 
\hline 
BJD\,--\,2400\,000 & RV & $\sigma_{\rm RV}$ & Bisector \\
(UTC)  & (km s$^{-1}$) & (km s$^{-1}$) & (km s$^{-1}$)\\ [0.5mm] \hline
\multicolumn{4}{l}{{\bf WASP-185:}}\\  
57191.68952 &   23.7717  &      0.0243 &        $-$0.0183\\
57193.69299 &   23.7864  &      0.0248 &        $-$0.0048 \\
57194.58975 &   23.8364  &      0.0150 &         0.0324 \\
57218.60712 &   23.8076  &      0.0267 &        $-$0.0193 \\
57221.56637 &   23.7937  &      0.0160 &         0.0179 \\
57412.87431 &   23.9926  &      0.0187 &         0.0396 \\
57487.74786 &   23.9708  &      0.0106 &         0.0601 \\
57488.68674 &   23.9470  &      0.0089 &         0.0051 \\
57591.52359 &   23.9950  &      0.0116 &         0.0061 \\
57599.55951 &   23.9768  &      0.0102 &         0.0201 \\
57809.83999 &   23.8754  &      0.0105 &        $-$0.0063\\
57815.86446 &   23.9621  &      0.0099 &         0.0093\\
57820.70709 &   23.8470  &      0.0114 &         0.0043\\
57901.58066 &   23.9342  &      0.0143 &        $-$0.0072\\
57905.75841 &   23.8140  &      0.0123 &         0.0364\\
57918.49287 &   23.9202  &      0.0122 &         0.0222\\
57933.62653 &   23.7907  &      0.0239 &        $-$0.0183\\
57951.54401 &   23.8020  &      0.0213 &         0.0090\\
57952.60960 &   23.7811  &      0.0148 &         0.0286\\
57990.48949 &   23.7948  &      0.0196 &        $-$0.0256\\
58311.55039 &   23.8551  &      0.0147 &         0.0610\\
58312.57675 &   23.8646  &      0.0148 &         0.0097\\
58324.57747 &   23.8505  &      0.0198 &        $-$0.0167\\
58357.50144 &   23.8219  &      0.0153 &        $-$0.0224\\ [0.5mm]
\hline
%
\multicolumn{4}{l}{{\bf WASP-192:}}\\  
57568.61900 &   15.5839 &       0.0340 &        $-$0.0655\\
58312.63770 &   16.2129 &       0.0304 &        $-$0.0336 \\
58329.60684 &   16.0837 &       0.0331 &        $-$0.0253 \\
58541.83740 &   15.6896 &       0.0533 &         0.1255 \\
58542.87037 &   16.2018 &       0.0453 &        $-$0.0570 \\
58543.80761 &   15.8034 &       0.0445 &        $-$0.0852 \\
58544.74536 &   15.6832 &       0.0514 &        $-$0.0766 \\
58544.88449 &   15.7419 &       0.0518 &         0.1486 \\
58545.72117 &   16.1702 &       0.1396 &         0.2143 \\
58576.82272 &   15.9405 &       0.0537 &         0.0655 \\
58576.88842 &   15.9702 &       0.0671 &        $-$0.0397 \\
58578.83595 &   15.5516 &       0.0611 &         0.0154 \\ [0.5mm]
\hline
\multicolumn{4}{l}{Bisector errors are twice RV errors} 
\end{tabular} 
\end{table}

\end{document}